\documentclass{article}
\usepackage{spconf}

\usepackage{amsmath,graphicx}
\usepackage[table]{xcolor}
\usepackage{booktabs}
\usepackage{amssymb,bm}
\usepackage{textcomp}
\usepackage{enumitem}
\usepackage{multirow}
\usepackage{adjustbox}
\usepackage{float}
\usepackage[colorlinks=true,linkcolor=black,anchorcolor=black,citecolor=black,filecolor=black,menucolor=black,runcolor=black,urlcolor=black]{hyperref}

\newcommand{\oldmindcf}{$\mathrm{NDCF_{0.01}^{min}}$}
\newcommand{\newmindcf}{$\mathrm{NDCF_{0.001}^{min}}$}

\title{A Multi Purpose and Large Scale Speech Corpus in Persian and English for Speaker and Speech Recognition: the DeepMine Database}

\name{{\em Hossein Zeinali\,$^{1,2}$, Luk\'a\v{s} Burget\,$^1$ and Jan ``Honza'' \v{C}ernock\'{y}\,$^1$}}

\address{$^1$ Brno University of Technology, Faculty of IT, IT4I Centre of Excellence, Czechia \\
	$^2$ Sharif DeepMine Ltd., Tehran, Iran
}

\begin{document}
\ninept
\maketitle
\begin{abstract}
DeepMine is a speech database in Persian and English designed to build and evaluate text-dependent, text-prompted, and text-independent speaker verification, as well as Persian speech recognition systems. It contains more than 1850 speakers and 540 thousand recordings overall, more than 480 hours of speech are transcribed. It is the first public large-scale speaker verification database in Persian, the largest public text-dependent and text-prompted speaker verification database in English, and the largest public evaluation dataset for text-independent speaker verification. It has a good coverage of age, gender, and accents. 
We provide several evaluation protocols for each part of the database to allow for research on different aspects of speaker verification. We also provide the results of several experiments that can be considered as baselines: HMM-based i-vectors for text-dependent speaker verification, and HMM-based as well as state-of-the-art deep neural network based ASR. We demonstrate that the database can serve for training robust ASR models.
\end{abstract}
\section{Introduction}
\label{sec.intro}

Nowadays deep learning techniques outperform the other conventional methods in most of the speech-related tasks. Training robust deep neural networks for each task depends on the availability of powerful processing GPUs, as well as standard and large scale datasets. In text-independent speaker verification, large-scale datasets are available, thanks to the NIST SRE evaluations and other data collection projects such as VoxCeleb~\cite{nagrani2017voxceleb}.

In text-dependent speaker recognition, experiments with end-to-end architectures conducted on large proprietary databases have demonstrated their superiority over traditional approaches~\cite{heigold2015end}. Yet, contrary to text-independent speaker recognition, text-dependent speaker recognition lacks large-scale publicly available databases. The two most well-known datasets are probably RSR2015~\cite{larcher2014text} and RedDots~\cite{lee2015reddots}. The former contains speech data collected from 300 individuals in a controlled manner, while the latter is used primarily for evaluation rather than training, due to its small number of speakers (only 64). Motivated by this lack of large-scale dataset for text-dependent speaker verification, we chose to proceed with the collection of the DeepMine dataset, which we expect to become a standard benchmark for the task.

Apart from speaker recognition, large amounts of training data are required also for training automatic speech recognition (ASR) systems. Such datasets should not only be large in size, they should also be characterized by high variability with respect to speakers, age and dialects. While several datasets with these properties are available for languages like English, Mandarin, French, this is not the case for several other languages, such as Persian. To this end, we proceeded with collecting a large-scale dataset, suitable for building robust ASR models in Persian.

The main goal of the DeepMine project was to collect speech from at least a few thousand speakers, enabling research and development of deep learning methods. The project started at the beginning of 2017, and after designing the database and the developing Android and server applications, the data collection began in the middle of 2017. The project finished at the end of 2018 and the cleaned-up and final version of the database was released at the beginning of 2019.
In~\cite{zeinali2018deepmine}, the running project and its data collection scenarios were described, alongside with some preliminary results and statistics. In this paper, we announce the final and cleaned-up version of the database, describe its different parts and provide various evaluation setups for each part. Finally, since the database was designed mainly for text-dependent speaker verification purposes, some baseline results are reported for this task on the official evaluation setups. Additional baseline results are also reported for Persian speech recognition. However, due to the space limitation in this paper, the baseline results are not reported for all the database parts and conditions. They will be defined and reported in the database technical documentation and in a future journal paper.

\section{Data Collection}
\label{sect:Data_Collection}

DeepMine is publicly available for everybody with a variety of licenses for different users. It was collected using crowdsourcing~\cite{zeinali2018deepmine}. The data collection was done using an Android application. Each respondent installed the application on his/her personal device and recorded several phrases in different sessions. The Android application did various checks on each utterance and if it passed all of them, the respondent was directed to the next phrase. For more information about data collection scenario, please refer to~\cite{zeinali2018deepmine}.

\subsection{Post-Processing}

In order to clean-up the database, the main post-processing step was to filter out problematic utterances. Possible problems include speaker word insertions (e.g. repeating some part of a phrase), deletions, substitutions, and involuntary disfluencies. To detect these, we implemented an alignment stage, similar to the second alignment stage in the LibriSpeech project~\cite{panayotov2015librispeech}. In this method, a custom decoding graph was generated for each phrase. The decoding graph allows for word skipping and word insertion in the phrase.

For text-dependent and text-prompted parts of the database, such errors are not allowed. Hence, any utterances with errors were removed from the enrollment and test lists. For the speech recognition part, a sub-part of the utterance which is correctly aligned to the corresponding transcription is kept. After the cleaning step, around 190 thousand utterances with full transcription and 10 thousand with sub-part alignment have remained in the database.

\subsection{Statistics}

After processing the database and removing problematic respondents and utterances, 1969 respondents\footnote{By ``respondent'', we mean a unique combination of speaker and mobile phone. Hence, one speaker with several different mobile devices is considered as several respondents.} remained in the database, with 1149 of them being male and 820 female. 297 of the respondents could not read English and have therefore read only the Persian prompts. About 13200 sessions were recorded by females and similarly, about 9500 sessions by males, i.e. women are over-represented in terms of sessions, even though their number is 17\:\% smaller than that of males. Other useful statistics related to the database are shown in Table~\ref{tbl.stats}.

\begin{table}[t]
	\renewcommand{\arraystretch}{1.1}
	\caption{\label{tbl.stats} Statistics of the DeepMine database.}
	\vspace{2mm}
	\centering{
		\setlength\tabcolsep{10pt}
		\begin{tabular}{ l c r }
			\toprule
			\midrule
			& & Count \\
			\midrule
			Finished sessions			        		& & 22,741 \\
			Finished sessions containing English	    & & 18,916 \\
			Recorded utterances						    & & 544,533 \\
			Unique speakers							    & & 1,858 \\
			Respondents with at least 5 utterances		& & 1,969 \\
			Respondents with at least 1 session		    & & 1,796 \\
			Respondents with at least 4 sessions		& & 1,104 \\
			Respondents with at least 8 sessions		& & 813 \\
			Respondents with at least 16 sessions		& & 607 \\
			Respondents with at least 60 sessions		& & 98 \\
			\midrule
			\bottomrule
		\end{tabular}
	}
\end{table}

The last status of the database, as well as other related and useful information about its availability can be found on its website, together with a limited number of samples\footnote{\url{http://data.deepmine.ir/en/}}.

\section{DeepMine Database Parts}
	
The DeepMine database consists of three parts. The first one contains fixed common phrases to perform text-dependent speaker verification. The second part consists of random sequences of words useful for text-prompted speaker verification, and the last part includes phrases with word- and phoneme-level transcription, useful for text-independent speaker verification using a random phrase (similar to Part4 of RedDots). This part can also serve for Persian ASR training. Each part is described in more details below. Table~\ref{tbl.phrase_counts} shows the number of unique phrases in each part of the database. For the English text-dependent part, the following phrases were selected from part1 of the RedDots database, hence the RedDots can be used as an additional training set for this part:
\begin{enumerate}
    \item ``My voice is my password.''
    \item ``OK Google.''
    \item ``Artificial intelligence is for real.''
    \item ``Actions speak louder than words.''
    \item ``There is no such thing as a free lunch.''
\end{enumerate}

\begin{table}[t]
	\renewcommand{\arraystretch}{1.1}
	\caption{\label{tbl.phrase_counts} Numbers of unique phrases in each phrase type. ``3-months/4-digits'' means a random subset of months/digits used as test phrase. ``Persian name and family'' means a sequence of several Persian full names.}
	\vspace{2mm}
	\centering{
		\setlength\tabcolsep{8pt}
		\begin{tabular}{ l c r c }
			\toprule
			\midrule
			& & Count & Part \\
			\midrule
			Persian text-dependent				& & 5           & 1 \\
			English text-dependent	        	& & 5           & 1 \\
			Persian months-name, 3-months		& & 1,320       & 2 \\
			Persian months-name, 12-months		& & 11,620      & 2 \\
			English digits, 4-digits			& & 5,040       & 2 \\
			English digits, 10-digits			& & 6,448       & 2 \\
			Persian name and family				& & 8,305       & 3 \\
			Persian transcribed phrases			& & 166,838     & 3 \\
			\midrule
			\bottomrule
		\end{tabular}
	}
\end{table}

\subsection{Part1 - Text-dependent (TD)}

This part contains a set of fixed phrases which are used to verify speakers in text-dependent mode. Each speaker utters 5 Persian phrases, and if the speaker can read English, 5 phrases selected from Part1 of the RedDots database are also recorded.

We have created three experimental setups with different numbers of speakers in the evaluation set. For each setup, speakers with more recording sessions are included in the evaluation set and the rest of the speakers are used for training in the background set (in the database, all background sets are basically training data). The rows in Table~\ref{tbl.td_sets} corresponds to the different experimental setups and shows the numbers of speakers in each set. Note that, for English, we have filtered the (Persian native) speakers by the ability to read English. Therefore, there are fewer speakers in each set for English than for Persian. There is a small ``dev'' set in each setup which can be used for parameter tuning to prevent over-tuning on the evaluation set.

\begin{table}[b]
	\renewcommand{\arraystretch}{1.1}
	\caption{\label{tbl.td_sets} Number of speakers in different evaluation sets of text-dependent part.}
	\vspace{2mm}
	\centering{
		\setlength\tabcolsep{4pt}
		\begin{tabular}{ l l c c c c c c c c }
			\toprule
			\midrule
						&	& & \multicolumn{3}{c}{Male} & & \multicolumn{3}{c}{Female} \\
							\cmidrule{4-6} \cmidrule{8-10}
			Lang		& Set name		& & back & dev & eval & & back & dev & eval \\
			\midrule
			\multirow{3}{*}{Persian}
			& 100-spk		& & 851 & 10 & 100 & & 607 & 10 & 100 \\
			& 200-spk		& & 741 & 20 & 200 & & 497 & 20 & 200 \\
			& 300-spk		& & 631 & 30 & 300 & & 387 & 30 & 300 \\
			\midrule
			\multirow{3}{*}{English}
			& 100-spk		& & 722 & 8  & 85  & & 510 & 8  & 88 \\
			& 200-spk		& & 620 & 18 & 175 & & 415 & 17 & 174 \\
			& 300-spk		& & 538 & 29 & 245 & & 318 & 27 & 261 \\
			\midrule
			\bottomrule
		\end{tabular}
	}
\end{table}

For each experimental setup, we have defined several official trial lists with different numbers of enrollment utterances per trial in order to investigate the effects of having different amounts of enrollment data. All trials in one trial list have the same number of enrollment utterances (3 to 6) and only one test utterance. All enrollment utterances in a trial are taken from different consecutive sessions and the test utterance is taken from yet another session. From all the setups and conditions, the 100-spk with 3-session enrollment (3-sess) is considered as the main evaluation condition. In Table~\ref{tbl.td_trials_count}, the number of trials for Persian 3-sess are shown for the different types of trial in the text-dependent speaker verification (SV). Note that for Imposter-Wrong (IW) trials (i.e. imposter speaker pronouncing wrong phrase), we merely create one wrong trial for each Imposter-Correct (IC) trial to limit the huge number of possible trials for this case. So, the number of trials for IC and IW cases are the same.

\begin{table}[t]
	\renewcommand{\arraystretch}{1.1}
	\caption{\label{tbl.td_trials_count} Numbers of trials in each different evaluation set for all possible trial types in text-dependent SV for 3-sess enrollment. TC: Target-Correct, TW: Target-Wrong, IC: Imposter-Correct, IW: Imposter-Wrong. K stands for thousand and M stands for million trials.}
	\vspace{2mm}
	\centering{
		\setlength\tabcolsep{2pt}
		\begin{adjustbox}{width=1.0\columnwidth}
		\begin{tabular}{ l c c c c c c c c c c}
			\toprule
			\midrule
					    & & \multicolumn{4}{c}{Male} & & \multicolumn{4}{c}{Female} \\
							\cmidrule{3-6} \cmidrule{8-11}
			Set name	& & TC & TW & IC & IW & & TC & TW & IC & IW \\
			\midrule
			100-spk		& & 409K & 1.63M & 30.7M & 30.7M & & 667K & 2.67M & 60.0M & 60.0M \\
			200-spk		& & 452K & 1.81M & 58.0M & 58.0M & & 792K & 3.17M & 126M  & 126M  \\
			300-spk		& & 480K & 1.92M & 85.6M & 85.6M & & 835K & 3.34M & 176M  & 176M  \\
			\midrule
			\bottomrule
		\end{tabular}
		\end{adjustbox}
	}
\end{table}
	
\subsection{Part2 - Text-prompted (TP)}

For this part, in each session, 3 random sequences of Persian month names are shown to the respondent in two modes: In the first mode, the sequence consists of all 12 months, which will be used for speaker enrollment. The second mode contains a sequence of 3 month names that will be used as a test utterance. In each 8 sessions received by a respondent from the server, there are 3 enrollment phrases of all 12 months (all in just one session), and $7 \times 3$ other test phrases, containing fewer words. For a respondent who can read English, 3 random sequences of English digits are also recorded in each session. In one of the sessions, these sequences contain all digits and the remaining ones contain only 4 digits.

Similar to the text-dependent case, three experimental setups with different number of speaker in the evaluation set are defined (corresponding to the rows in Table~\ref{tbl.tp_trials_count}). However, different strategy is used for defining trials: Depending on the enrollment condition (1- to 3-sess), trials are enrolled on utterances of all words from 1 to 3 different sessions (i.e. 3 to 9 utterances). Further, we consider two conditions for test utterances: \textbf{seq} test utterance with only 3 or 4 words and \textbf{full} test utterances with all words (i.e. same words as in enrollment but in different order). From all setups an all conditions, the 100-spk with 1-session enrolment (1-sess) is considered as the main evaluation condition for the text-prompted case. In Table~\ref{tbl.tp_trials_count}, the numbers of trials (sum for both seq and full conditions) for Persian 1-sess are shown for the different types of trials in the text-prompted SV. Again, we just create one IW trial for each IC trial.

\vspace{-1mm}
\begin{table}[b]
	\renewcommand{\arraystretch}{1.1}
	\caption{\label{tbl.tp_trials_count} Numbers of trials in each different evaluation sets for all possible trial types in text-prompted SV for 1-sess enrollment.}
	\vspace{2mm}
	\centering{
		\setlength\tabcolsep{2pt}
		\begin{adjustbox}{width=1.0\columnwidth}
		\begin{tabular}{ l c c c c c c c c c c}
			\toprule
			\midrule
					    & & \multicolumn{4}{c}{Male} & & \multicolumn{4}{c}{Female} \\
							\cmidrule{3-6} \cmidrule{8-11}
			Set name	& & TC & TW & IC & IW & & TC & TW & IC & IW \\
			\midrule
			100-spk		& & 98.5K & 394K & 7.61M & 7.61M & & 155K & 621K & 14.2M & 14.2M \\
			200-spk		& & 111K  & 446K & 15.2M & 15.2M & & 187K & 749K & 30.5M & 30.5M \\
			300-spk		& & 119K  & 474K & 22.5M & 22.5M & & 199K & 798K & 43.9M & 43.9M \\
			\midrule
			\bottomrule
		\end{tabular}
		\end{adjustbox}
	}
\end{table}

\subsection{Part3 - Text-independent (TI)}

In this part, 8 Persian phrases that have already been transcribed on the phone level are displayed to the respondent. These phrases are chosen mostly from news and Persian Wikipedia. If the respondent is unable to read English, instead of 5 fixed phrases and 3 random digit strings, 8 other Persian phrases are also prompted to the respondent to have exactly 24 phrases in each recording session.

This part can be useful at least for three potential applications. First, it can be used for text-independent speaker verification. The second application of this part (same as Part4 of RedDots) is text-prompted speaker verification using random text (instead of a random sequence of words). Finally, the third application is large vocabulary speech recognition in Persian (explained in the next sub-section).

Based on the recording sessions, we created two experimental setups for speaker verification. In the first one, respondents with at least 17 recording sessions are included to the evaluation set, respondents with 16 sessions to the development and the rest of respondents to the background set (can be used as training data). In the second setup, respondents with at least 8 sessions are included to the evaluation set, respondents with 6 or 7 sessions to the development and the rest of respondents to the background set. Table~\ref{tbl.ti_sets} shows numbers of speakers in each set of the database for text-independent SV case.

\begin{table}[t]
	\renewcommand{\arraystretch}{1.1}
	\caption{\label{tbl.ti_sets} Different evaluation sets of text-independent part of the DeepMine database.}
	\vspace{2mm}
	\centering{
		\setlength\tabcolsep{6pt}
		\begin{tabular}{ l c c c c c c c c }
			\toprule
			\midrule
						& & \multicolumn{3}{c}{Male} & & \multicolumn{3}{c}{Female} \\
							\cmidrule{3-5} \cmidrule{7-9}
			Set name	& & back & dev & eval & & back & dev & eval \\
			\midrule
			438-spk	    & & 689 & 91 & 181 & & 376 & 85 & 257 \\
			798-spk	    & & 540 & 59 & 365 & & 220 & 65 & 433 \\
			\midrule
			\bottomrule
		\end{tabular}
	}
\end{table}

For text-independent SV, we have considered 4 scenarios for enrollment and 4 scenarios for test. The speaker can be enrolled using utterances from 1, 2 or 3 consecutive sessions (1sess to 3sess) or using 8 utterances from 8 different sessions. The test speech can be one utterance (1utt) for short duration scenario or all utterances in one session (1sess) for long duration case. In addition, test speech can be selected from 5 English phrases for cross-language testing (enrollment using Persian utterances and test using English utterances). From all setups, 1sess-1utt and 1sess-1sess for 438-spk set are considered as the main evaluation setups for text-independent case. Table~\ref{tbl.ti_trials_count} shows number of trials for these setups.

\begin{table}[b]
	\renewcommand{\arraystretch}{1.1}
	\caption{\label{tbl.ti_trials_count} Numbers of trials in each different evaluation sets for all possible trial types in text-dependent SV for 3-sess enrollment. ``cross'' means enrollment using Persian utterances and test using 5 English utterances. K stands for thousand and M stands for million trials.}
	\vspace{2mm}
	\centering{
		\setlength\tabcolsep{4pt}
		\begin{tabular}{ l c c c c c c }
			\toprule
			\midrule
					            & & \multicolumn{2}{c}{Male} & & \multicolumn{2}{c}{Female} \\
						    	\cmidrule{3-4} \cmidrule{6-7}
			Set name    	    & & Target & Imposter & & Target & Imposter \\
			\midrule
			1sess-1utt		    & & 537K & 61.7M & & 1013K & 193.1M \\
			1sess-1utt-cross	& & 459K & 51.3M & & 890K  & 164.4M \\
			1sess-1sess		    & & 135K & 15.6M & & 254K  & 48.7M  \\
			1sess-1sess-cross	& & 229K & 25.6M & & 445K  & 82.2M  \\
			\midrule
			\bottomrule
		\end{tabular}
	}
\end{table}

For text-prompted SV with random text, the same setup as text-independent case together with corresponding utterance transcriptions can be used.

\subsection{Part3 - Speech Recognition}

As explained before, Part3 of the DeepMine database can be used for Persian read speech recognition. There are only a few databases for speech recognition in Persian~\cite{bijankhan1994farsdat, bijankhan1994tfarsdat}. Hence, this part can at least partly address this problem and enable robust speech recognition applications in Persian. Additionally, it can be used for speaker recognition applications, such as training deep neural networks (DNNs) for extracting bottleneck features~\cite{zeinali2016csl}, or for collecting sufficient statistics using DNNs for i-vector training.

We have randomly selected 50 speakers (25 for each gender) from the all speakers in the database which have net speech (without silence parts) between 25 minutes to 50 minutes as test speakers. For each speaker, the utterances in the first 5 sessions are included to (small) test-set and the other utterances of test speakers are considered as a large-test-set. The remaining utterances of the other speakers are included in the training set. The test-set, large-test-set and train-set contain 5.9, 28.5 and 450 hours of speech respectively.

There are about 8300 utterances in Part3 which contain only Persian full names (i.e. first and family name pairs). Each phrase consists of several full names and their phoneme transcriptions were extracted automatically using a trained Grapheme-to-Phoneme (G2P). These utterances can be used to evaluate the performance of a systems for name recognition, which is usually more difficult than the normal speech recognition because of the lack of a reliable language model.

\section{Experiments and Results}
\label{sec:results}

Due to the space limitation, we present results only for the Persian text-dependent speaker verification and speech recognition.

\subsection{Speaker Verification Experiments}

We conducted an experiment on text-dependent speaker verification part of the database, using the i-vector based method proposed in \cite{zeinali2016deep, zeinali2016trans} and applied it to the Persian portion of Part1. In this experiment, 20-dimensional MFCC features along with first and second derivatives are extracted from 16\,kHz signals using HTK~\cite{young1997htk} with 25\,ms Hamming windowed frames with 15\,ms overlap.

The reported results are obtained with a 400-dimensional gender independent i-vector based system. The i-vectors are first length-normalized and are further normalized using phrase- and gender-dependent Regularized Within-Class Covariance Normalization (RWCCN)~\cite{zeinali2016trans}. Cosine distance is used to obtain speaker verification scores and phrase- and gender-dependent s-norm is used for normalizing the scores. For aligning speech frames to Gaussian components, monophone HMMs with 3 states and 8 Gaussian components in each state are used~\cite{zeinali2016trans}. We only model the phonemes which appear in the 5 Persian text-dependent phrases.

For speaker verification experiments, the results were reported in terms of Equal Error Rate (EER) and Normalized Detection Cost Function as defined for NIST SRE08 (\oldmindcf) and NIST SRE10 (\newmindcf). As shown in Table~\ref{tbl.trial_types}, in text-dependent SV there are 4 types of trials: {\em Target-Correct} and {\em Imposter-Correct} refer to trials when the pass-phrase is uttered correctly by target and imposter speakers respectively, and in same manner, {\em Target-Wrong} and {\em Imposter-Wrong} refer to trials when speakers uttered a wrong pass-phrase. In this paper, only the correct trials (i.e. Target-Correct as target trials vs Imposter-Correct as non-target trials) are considered for evaluating systems as it has been proved that these are the most challenging trials in text-dependent SV~\cite{zeinali2016csl, zeinali2016ivector}.

\begin{table}[t]
  \renewcommand{\arraystretch}{1.2}
  \caption{\label{tbl.trial_types} Trial types in text-dependent speaker verification~\cite{larcher2014text}.}
  \vspace{2mm}
  \centerline
  {
    \begin{tabular}{l | c c}
      \toprule
       & \bfseries{Target Speaker} & \bfseries{Imposter Speaker} \\
      \midrule
      \bfseries{Correct Pass-Phrase}	& Target-Correct	& Imposter-Correct \\
      \bfseries{Wrong Pass-Phrase}		& Target-Wrong		& Imposter-Wrong \\
      \bottomrule
    \end{tabular}
  }
\end{table}

\begin{table*}[tb!]
	\renewcommand{\arraystretch}{1.1}
	\caption{\label{tbl.text_dependent_results} i-vector/HMM results on text-dependent part (i.e. Part1) on 100-spk, 3-sess setup of Persian language. The results are only for correct trials (i.e. Target-Correct vs Imposter-Corrects). The first column shows filtering of the trials in target:imposter format. The first Y letter on each side of the colon shows the condition where the same mobile brand and the second Y shows exactly the same device models were used for recording the enrollment and test utterances.}
	\vspace{2mm}
	\centerline
	{
		\setlength\tabcolsep{4.5pt}
		\begin{tabular}{l | c c c c c | c c c c c }
			\toprule
			\midrule
			       & \multicolumn{5}{c}{\textbf{Male}} & \multicolumn{5}{c}{\textbf{Female}} \\
			       \cmidrule{2-11}
			Filter & \# Target & \# Imposter & EER [\%] & \oldmindcf & \newmindcf & \# Target & \# Imposter & EER [\%] & \oldmindcf & \newmindcf \\
			\midrule
			-- -- : -- --	& 408951 & 30742420 & 1.54 & 0.189 & 0.633 & 666946 & 59953740	& 3.01  & 0.163 & 0.318 \\
			NN : YY			& 55950	 & 233560	& 4.61 & 0.297 & 0.432 & 71459 	& 434464	& 15.54 & 0.562 & 0.701 \\
			YY : -- --		& 346799 & 30742420	& 1.09 & 0.159 & 0.568 & 560827	& 59953740	& 1.28	& 0.107 & 0.247 \\
			YY : YY			& 346799 & 233560	& 1.48 & 0.098 & 0.153 & 560827	& 434464	& 1.79	& 0.138 & 0.217 \\
			YY : YN			& 346799 & 8216620	& 1.13 & 0.084 & 0.142 & 560827	& 20902608	& 1.33	& 0.097 & 0.158 \\
			Y -- : -- --	& 353001 & 30742420	& 1.11 & 0.163 & 0.578 & 595487	& 59953740	& 1.49	& 0.123 & 0.271 \\
			-- N : Y --		& 62152 & 8450180	& 3.49 & 0.258 & 0.408 & 106119	& 21337072	& 10.45 & 0.439 & 0.590 \\
			\midrule
			\bottomrule
		\end{tabular}
	}
	\vspace{-1mm}
\end{table*}

Table~\ref{tbl.text_dependent_results} shows the results of text-dependent experiments using Persian 100-spk and 3-sess setup. For filtering trials, the respondents' mobile brand and model were used in this experiment. In the table, the first two letters in the filter notation relate to the target trials and the second two letters (i.e. right side of the colon) relate for non-target trials. For target trials, the first Y means the enrolment and test utterances were recorded using a device with the same brand by the target speaker. The second Y letter means both recordings were done using exactly the same device model. Similarly, the first Y for non-target trials means that the devices of target and imposter speakers are from the same brand (i.e. manufacturer). The second Y means that, in addition to the same brand, both devices have the same model. So, the most difficult target trials are ``NN'', where the speaker has used different a device at the test time. In the same manner, the most difficult non-target trials which should be rejected by the system are ``YY'' where the imposter speaker has used the same device model as the target speaker (note that it does not mean physically the same device because each speaker participated in the project using a personal mobile device). Hence, the similarity in the recording channel makes rejection more difficult.

The first row in Table~\ref{tbl.text_dependent_results} shows the results for all trials. By comparing the results with the best published results on RSR2015 and RedDots~\cite{zeinali2016trans, zeinali2016csl, zeinali2016ivector}, it is clear that the DeepMine database is more challenging than both RSR2015 and RedDots databases. For RSR2015, the same i-vector/HMM-based method with both RWCCN and s-norm has achieved EER less than 0.3\:\% for both genders (Table VI in~\cite{zeinali2016trans}). The conventional Relevance MAP adaptation with HMM alignment without applying any channel-compensation techniques (i.e. without applying RWCCN and s-norm due to the lack of suitable training data) on RedDots Part1 for the male has achieved EER around 1.5\:\% (Table XI in~\cite{zeinali2016trans}). It is worth noting that EERs for DeepMine database without any channel-compensation techniques are 2.1 and 3.7\:\% for males and females respectively.

One interesting advantage of the DeepMine database compared to both RSR2015 and RedDots is having several target speakers with more than one mobile device. This is allows us to analyse the effects of channel compensation methods. The second row in Table~\ref{tbl.text_dependent_results} corresponds to the most difficult trials where the target trials come from mobile devices with different models while imposter trials come from the same device models. It is clear that severe degradation was caused by this kind of channel effects (i.e. decreasing within-speaker similarities while increasing between-speaker similarities), especially for females.

The results in the third row show the condition when target speakers at the test time use exactly the same device that was used for enrollment. Comparing this row with the results in the first row proves how much improvement can be achieved when exactly the same device is used by the target speaker.

The results in the fourth row show the condition when imposter speakers also use the same device model at test time to fool the system. So, in this case, there is no device mismatch in all trials. By comparing the results with the third row, we can see how much degradation is caused if we only consider the non-target trials with the same device.

The fifth row shows similar results when the imposter speakers use device of the same brand as the target speaker but with a different model. Surprisingly, in this case, the degradation is negligible and it means that mobiles from a specific brand (manufacturer) have different recording channel properties. 

The degraded female results in the sixth row as compared to the third row show the effect of using a different device model from the same brand for target trials. For males, the filters brings almost the same subsets of trials, which explains the very similar results in this case. 

Looking at the first two and the last row of Table~\ref{tbl.text_dependent_results}, one can notice the significantly worse performance obtained for the female trials as compared to males. Note that these three rows include target trials where the devices used for  enrollment do not necessarily match the devices used for recording test utterances. On the other hand, in rows 3 to 6, which exclude such mismatched trials, the performance for males and females is comparable. This suggest that the degraded results for females are caused by some problematic trials with device mismatch. The exact reason for this degradation is so far unclear and needs a further investigation.

In the last row of the table, the condition of the second row is relaxed: the target device should have different model possibly from the same brand and imposter device only needs to be from the same brand. In this case, as was expected, the performance degradation is smaller than in the second row.

\subsection{Speech Recognition Experiments}

In addition to speaker verification, we present several speech recognition experiments on Part3. The experiments were performed with the Kaldi toolkit~\cite{povey2011kaldi}. For training HMM-based MonoPhone model, only 20 thousands of shortest utterances are used and for other models the whole training data is used. The DNN based acoustic model is a time-delay DNN with low-rank factorized layers and skip connections without i-vector adaptation (a modified network from one of the best performing LibriSpeech recipes). The network is shown in Table~\ref{tbl.ftdnn}: there are 16 F-TDNN layers, with dimension 1536 and linear bottleneck layers of dimension 256. The acoustic model is trained for 10 epochs using lattice-free maximum mutual information (LF-MMI) with cross-entropy regularization~\cite{povey2016purely}. Re-scoring is done using a pruned trigram language model and the size of the dictionary is around 90,000 words. 

Table~\ref{tbl.ars_results} shows the results in terms of word error rate (WER) for different evaluated methods. As can be seen, the created database can be used to train well performing and practically usable Persian ASR models. 

\begin{table}[t]
	\renewcommand{\arraystretch}{1.1}
	\caption{\label{tbl.ftdnn} Factorized TDNN architecture for acoustic modeling using Kaldi toolkit~\cite{povey2011kaldi}. Note that Batch-Norm applied after each ReLU is omitted for brevity.}
	\vspace{2mm}
	\centerline
	{
	    \setlength\tabcolsep{3pt}
	    \begin{adjustbox}{width=1.0\columnwidth}
		\begin{tabular}{l c c c c c c}
			\toprule
			\midrule
		    Layer & Layer type & Context & Context & Skip conn. & Size & Inner \\
		          &            & factor1 & factor2 & from layer &      & size \\
		    \midrule
		    1  & TDNN-ReLU   & t      &        &           & 1536 &     \\
		    2  & F-TDNN-ReLU & t-1, t & t, t+1 & 0 (input) & 1536 & 256 \\
		    3  & F-TDNN-ReLU & t-1, t & t, t+1 & 1         & 1536 & 256 \\
		    4  & F-TDNN-ReLU & t-1, t & t, t+1 & 2         & 1536 & 256 \\
		    5  & F-TDNN-ReLU & t      & t      & 3         & 1536 & 256 \\
		    6  & F-TDNN-ReLU & t-3, t & t, t+3 & 4         & 1536 & 256 \\
		    7  & F-TDNN-ReLU & t-3, t & t, t+3 & 5         & 1536 & 256 \\
		    8  & F-TDNN-ReLU & t-3, t & t, t+3 & 6         & 1536 & 256 \\
		    9  & F-TDNN-ReLU & t-3, t & t, t+3 & 7         & 1536 & 256 \\
		    10 & F-TDNN-ReLU & t-3, t & t, t+3 & 8         & 1536 & 256 \\
		    11 & F-TDNN-ReLU & t-3, t & t, t+3 & 9         & 1536 & 256 \\
		    12 & F-TDNN-ReLU & t-3, t & t, t+3 & 10        & 1536 & 256 \\
		    13 & F-TDNN-ReLU & t-3, t & t, t+3 & 11        & 1536 & 256 \\
		    14 & F-TDNN-ReLU & t-3, t & t, t+3 & 12        & 1536 & 256 \\
		    15 & F-TDNN-ReLU & t-3, t & t, t+3 & 13        & 1536 & 256 \\
		    16 & F-TDNN-ReLU & t-3, t & t, t+3 & 14        & 1536 & 256 \\
		    17 & F-TDNN-ReLU & t-3, t & t, t+3 & 15        & 1536 & 256 \\
		    \midrule
		    18 & Linear      &        &        &           & 256  &     \\
		    19 & Dense-ReLU-Linear  & &        &           & 256  & 1536 \\
		    20 & Dense       &        &        &           & \# Tar. &     \\
		    \midrule
            \bottomrule
            \end{tabular}
	    \end{adjustbox}
	}
\end{table}

\begin{table}[tb!]
	\renewcommand{\arraystretch}{1.1}
	\caption{\label{tbl.ars_results} WER [\%] obtained with different ASR systemss trained on DeepMine database. Test-set has 5.9 hours of speech and Large-Test-set contains  28.5 hours of speech.}
	\vspace{2mm}
	\centering{
		\setlength\tabcolsep{4pt}
		\begin{tabular}{ l | c | c }
			\toprule
			\midrule
			& Test-set & Large-Test-set \\
			\midrule
			MonoPhone			        			& 41.31 & 40.70 \\
			TriPhone + Deltas + Delta-Deltas	   	& 17.30 & 16.50 \\
			TriPhone + LDA + MLLT					& 14.08 & 13.32 \\
			TriPhone + LDA + MLLT + SAT				& 11.52 & 10.95 \\
			\midrule
			Kaldi F-TDNN							&  4.44 &  4.09 \\
			\midrule
			\bottomrule
		\end{tabular}
	}
\end{table}

\section{Conclusions}
\label{sec.conc}

In this paper, we have described the final version of a large speech corpus, the DeepMine database. It has been collected using crowdsourcing and, according to the best of our knowledge, it is the largest public text-dependent and text-prompted speaker verification database in two languages: Persian and English. In addition, it is the largest text-independent speaker verification evaluation database, making it suitable to robustly evaluate state-of-the-art methods on different conditions. Alongside these appealing properties, it comes with phone-level transcription, making it suitable to train deep neural network models for Persian speech recognition.

We provided several evaluation protocols for each part of the database. The protocols allow researchers to investigate the performance of different methods in various scenarios and study the effects of channels, duration and phrase text on the performance. We also provide two test sets for speech recognition: One normal test set with a few minutes of speech for each speaker and one large test set with more (30 minutes on average) speech that can be used for any speaker adaptation method.

As baseline results, we reported the performance of an i-vector/HMM based method on Persian text-dependent part. Moreover, we conducted speech recognition experiments using conventional HMM-based methods, as well as state-of-the-art deep neural network based method using Kaldi toolkit with promising performance. Text-dependent results have shown that the DeepMine database is more challenging than RSR2015 and RedDots databases.

\section{Acknowledgments}
The data collection project was mainly supported by Sharif DeepMine company. The work on the paper was supported by Czech National Science Foundation (GACR) project "NEUREM3" No. 19-26934X and the National Programme of Sustainability (NPU II) project "IT4Innovations excellence in science - LQ1602".

\vspace{-1mm}
\bibliographystyle{IEEEbib}
\bibliography{Speaker}
	
\end{document}